\documentclass[%
 reprint,
 amsmath,amssymb,
 aps,
 superscriptaddress
]{revtex4-1}

\usepackage{graphicx}
\usepackage{dcolumn}
\usepackage{bm}
\usepackage{hyperref}

\begin{document}

\preprint{APS/123-QED}

\title{Approaching a realistic force balance in geodynamo simulations}

\author{Rakesh Yadav}
\affiliation{Harvard-Smithsonian Center for Astrophysics, 60 Garden Street, 02138 Cambridge, USA}
\affiliation{Max-Planck-Institut f\"{u}r Sonnensystemforschung, Justus-von-Liebig-Weg 3, 37077 G\"{o}ttingen, Germany}
\author{Thomas Gastine}
\affiliation{Max-Planck-Institut f\"{u}r Sonnensystemforschung, Justus-von-Liebig-Weg 3, 37077 G\"{o}ttingen, Germany}
\affiliation{Institut de Physique du Globe de Paris, Sorbonne Paris Cit\'e, Universit\'e
Paris-Diderot, UMR 7154 CNRS, 1 rue Jussieu, F-75005 Paris, France}
\author{Ulrich Christensen}
\affiliation{Max-Planck-Institut f\"{u}r Sonnensystemforschung, Justus-von-Liebig-Weg 3, 37077 G\"{o}ttingen, Germany}
\author{Scott J. Wolk}
\affiliation{Harvard-Smithsonian Center for Astrophysics, 60 Garden Street, 02138 Cambridge, USA}
\author{Katja Poppenhaeger}
\affiliation{Astrophysics Research Center, Queen's University Belfast, BT7 1NN Belfast, United Kingdom}
\affiliation{Harvard-Smithsonian Center for Astrophysics, 60 Garden Street, 02138 Cambridge, USA}
\date{\today}

\begin{abstract}
{\bf Paper published in the {\em Proceedings of the National Academy of Sciences (PNAS)}} \\
Earth sustains its magnetic field by a dynamo process driven by convection in the liquid outer core. Geodynamo simulations have been successful in reproducing many observed properties of the geomagnetic field. However, while theoretical considerations suggest that flow in the core is governed by a balance between Lorentz force, rotational force and buoyancy (called MAC balance for Magnetic, Archimedean, Coriolis) with only minute roles for viscous and inertial forces, dynamo simulations must employ viscosity  values that are many orders of magnitude larger than in the core due to computational constraints. In typical geodynamo models viscous and inertial forces are not much smaller than the Coriolis force and the Lorentz force plays a sub-dominant role. This has led to conclusions that these simulations are viscously controlled and do not represent the physics of the geodynamo. Here we show by a direct analysis of the relevant forces that a MAC balance can be achieved when the viscosity is reduced to values close to the current practical limit. Lorentz force, buoyancy and the uncompensated (by pressure) part of the Coriolis force are of very similar strength, whereas viscous and inertia are smaller by a factor of at least 20 in the bulk of the fluid volume. Compared to non-magnetic convection at otherwise identical parameters, the dynamo flow is of larger scale, less invariant parallel to the rotation axis (less geostrophic) and convection transports twice as much heat, all of which is expected when the Lorentz force strongly influences the convection properties. 

\end{abstract}

\maketitle

\section{Introduction}

Sustained magnetism in astrophysical objects is due to the dynamo mechanism which relies on the generation of electrical currents by fluid motion~\cite{brandenburg2005}. The secular cooling of the Earth's interior and the release of light elements at the boundary of the solid inner core provide buoyancy sources that drive convection, leading to the generation of electrical currents~\cite{roberts2013}. It has been more than two decades since the idea of modeling the geomagnetic field using computer simulations was successfully demonstrated~\cite{glatzmaier1995a,glatzmaier1995b}. These pioneering simulations were able to reproduce the dipole dominant nature of the geomagnetic field and showed reversals of the geomagnetic dipole. Since then computer simulations have become a primary tool for studying the properties of the geomagnetic field~\cite{christensen2004,gubbins2011,aubert2013, olson2014,aubert2014}.

The range of flow length scales present in the liquid outer core is enormous due to the very small viscosity of the fluid. To model this aspect in geodynamo simulations one would require tremendous  computing power that is not available even in the foreseeable future. Therefore, all geodynamo simulations must use unrealistically large viscosity to reduce the level of turbulence. One quantity that epitomizes this discrepancy is the Ekman number $E=\nu\Omega^{-1}D^{-2}$ ($\nu$ is the viscosity, $\Omega$ is the Earth's rotation rate, and $D$ is the thickness of the liquid outer core) which roughly quantifies the ratio of the viscous force ${F}_V$ and the Coriolis force ${F}_C$. The Ekman number is about $10^{-15}$ in the core while simulations typically use $10^{-4}$ \cite{roberts2013}. 

The Coriolis force tends to suppress changes of the flow in the direction of the rotation axis, i.e., makes the flow nearly geostrophic~\cite{proudman1916, taylor1923}. This is known as the ``Proudman-Taylor constraint" (PTC). Because the boundary of the fluid core is inclined relative to the direction of rotation (except at the poles), convective motions cannot be purely geostrophic and therefore the PTC impedes convection~\cite{greenspan1968}. In the absence of a magnetic field viscous force or inertial force $F_I$ must compensate the part of the Coriolis force that cannot be balanced by the pressure force $F_P$. $F_V$ or $F_I$ may still be significantly smaller than the Coriolis force. For example, at the onset of non-magnetic convection in a sphere, $F_V$ is smaller than $F_C$ by a factor $E^{1/3}$. Nonetheless, it is of the same order as $|{\bf F}_C + {\bf F}_P|$ and plays a key role in the force balance. The buoyancy force $F_A$ (Archimedean) is comparable to $F_V$ and the state can be referred to as being in a VAC-balance (Viscous, Archimedean, Coriolis) \cite{king2013b}.

In the Earth's core, the buoyancy force and the Lorentz force $F_L$ due to the geomagnetic field are expected to be comparable to the Coriolic force \cite{chandrasekhar1954, stevenson1979,starchenko2002, roberts2013}. This state is commonly referred to as "MAC" state. Here, the dynamo presumably selects a magnetic field that leads to an efficient relaxation of the PTC. This is expected to occur at $\Lambda\approx\mathcal{O}(1)$, where the Elsasser number is $\Lambda = {B^2}(\rho\mu\lambda\Omega)^{-1}$ ($B$ is mean magnetic field, $\rho$ is density, $\mu$ is magnetic permeability, $\lambda$ is magnetic diffusivity) \cite{chandrasekhar1954, stevenson1979}.  Note that here we use the term MAC-balance in the sense that $F_L$ and $F_A$ are of the same order as the uncompensated Coriolis force $|{\bf F}_C + {\bf F}_P|$, not necessarily the total Coriolis force.

Although a MAC state has long been expected from theoretical considerations, its existence in  geodynamo simulations has not been demonstrated so far. A recent study of geodynamo models at an Ekman number of $10^{-4}$ explicitly calculated the value of the various forces \cite{soderlund2012}. The authors show that the viscous force was actually comparable to the other forces. Furthermore, the analysis of convection properties suggested that a VAC state exists in contemporary geodynamo simulations rather than a MAC state \cite{king2013b}.  The presence of a VAC state promotes the idea that cost-efficient simulations might produce geodynamo-like features for the wrong reasons \cite{roberts2013}. A natural question then arises: How small should the viscosity be for a MAC state to appear? Due to the very nature of this question a detailed parameter study is called for that systematically explores the parameter regime of geodynamo simulations.

\section{Methods}
We carry out a detailed study of geodynamo models where we analyze data from our recent study \cite{yadav2016} and carry out new simulations at more extreme values of the control parameters. The basic setup is geodynamo-like and we consider a spherical shell where the ratio of the inner ($r_i$) and the outer ($r_o$) radius is 0.35. The thickness $D$ of the shell is given by $r_o-r_i$. The convection in the shell is driven by a superadiabatic temperature contrast $\Delta T$ across the two boundaries. The shell rotates along the $\hat z$ axis with an angular frequency $\Omega$. We work with non-dimensional equations and we use $D$ as standard length scale, $D^2/\nu$ as time scale, $\Delta T$ as  temperature scale, and $\sqrt{\rho\mu\lambda\Omega}$ as magnetic field scale.

We employ the Boussinesq approximation and the equations governing the velocity $\mathbf{u}$, magnetic field $\mathbf{B}$, and temperature perturbation $T$ are:
\begin{gather}
E\left(\frac{\partial\mathbf{u}}{\partial t}+\mathbf{u\cdot\nabla\mathbf{u}}\right)+2\hat{z}\times\mathbf{u}= -\nabla P + \frac{Ra\,E}{P_r}\,{g(r)\,T\,\hat{\bf r}} \nonumber \\
 +{\frac{1}{P_m}}(\nabla\times\mathbf{B})\times\mathbf{B}+E\nabla^{2}\mathbf{u},  \label{eq:MHD_vel} \\
\nabla \cdot \mathbf{u}  =  0, \\
\frac{\partial T}{\partial t}+\mathbf{u\cdot\nabla}T  = \frac{1}{P_r}\nabla^{2}T, \\
\frac{\partial\mathbf{B}}{\partial t}  =  \nabla\times(\mathbf{u}\times\mathbf{B})+\frac{1}{P_m}\nabla^{2}\mathbf{B}, \label{eq:MHD_mag} \\
\nabla \cdot \mathbf{B}  =  0,  \label{eq:div_B_0}
\end{gather}
where $g(r)$ is the gravity that varies as $r/r_o$, and $P$ is the pressure. The  control parameters that govern the system are:
\begin{gather}
\text{Prandtl number } \,\,\,\, P_r=\frac{\nu}{\kappa},  \\
\text{magnetic Prandtl number }  \,\,\,\, P_m=\frac{\nu}{\lambda},  \\
\text{Rayleigh number }  \,\,\,\, Ra=\frac{\alpha\,g_o\,D^3\Delta T}{\nu\,\kappa},
\end{gather}
where $\alpha$ is the thermal expansivity, and $g_o$ is the gravity at the outer boundary, and $\kappa$ is the thermal diffusivity.

Both boundaries have fixed temperature, are no-slip, and are electrically insulating. The open-source code MagIC (available at \href{https://github.com/magic-sph/magic}{\tt www.github.com/magic-sph/magic}) is used to simulate the models \cite{wicht2002}. The code uses spherical harmonic decomposition in latitude and longitude and Chebyshev polynomials in the radial direction. MagIC uses the  SHTns library \cite{schaeffer2013} to efficiently calculate the spherical harmonic transforms. Since we employ non-dimensional equations, the relative influence of viscosity is mainly expressed by the value of the Ekman number. To explore the effect of the magnetic field we perform hydrodynamic (HD) simulations, i.e. without a magnetic field, in parallel to the dynamo models. 

The results of simulations with $E=10^{-4},\,10^{-5}$ are taken from our earlier study \cite{yadav2016} and are extended here to runs at $E=10^{-6}$. In all of our simulations, the fluid Prandtl number $P_r$ is unity. The magnetic Prandtl number $P_m$ is also unity for cases with $E=10^{-4}$ and $E=10^{-5}$. At $E=10^{-6}$, we ran five dynamo simulations with $P_m$ of 2, 1, 0.5, 0.5, and 0.4 (in order of increasing $Ra$). To reduce the time spent in calculating the transient stages for $E=10^{-6}$ simulation with   highest $Ra$ we use a scaled $E=10^{-5}$ dynamo simulation as initial condition. The scaling factors for magnetic field and velocity are calculated using the scaling laws by Christensen \& Aubert \cite{christensen2006}. Furthermore, the other $E=10^{-6}$ simulations at lower $Ra$ use an initial condition from a higher $Ra$ case. Data tables that contain useful globally-averaged quantities, grid resolutions, and simulation run-time are provided as online supplementary material.

\begin{figure*}
\centering
\includegraphics[width=0.8\linewidth]{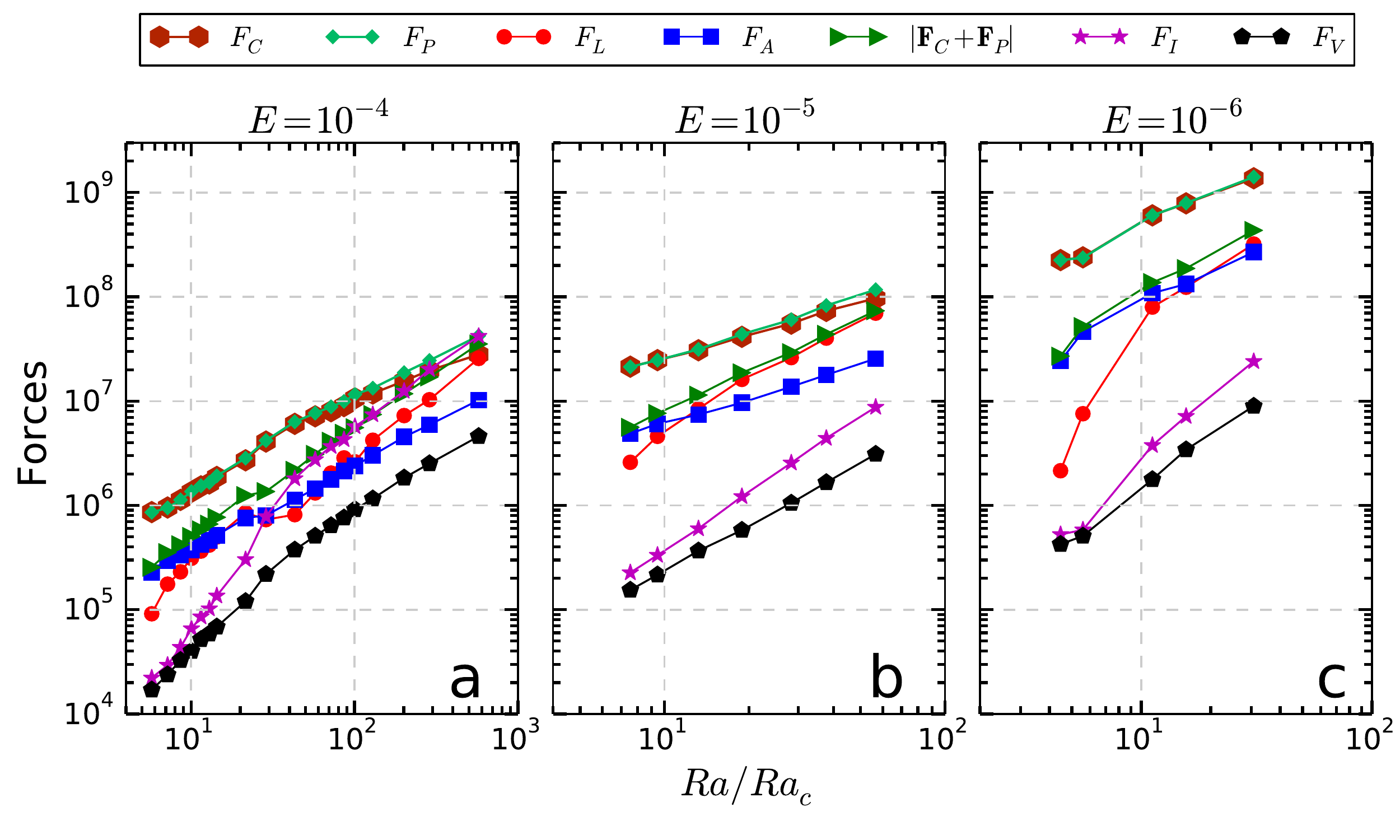} \\
\includegraphics[width=0.8\linewidth]{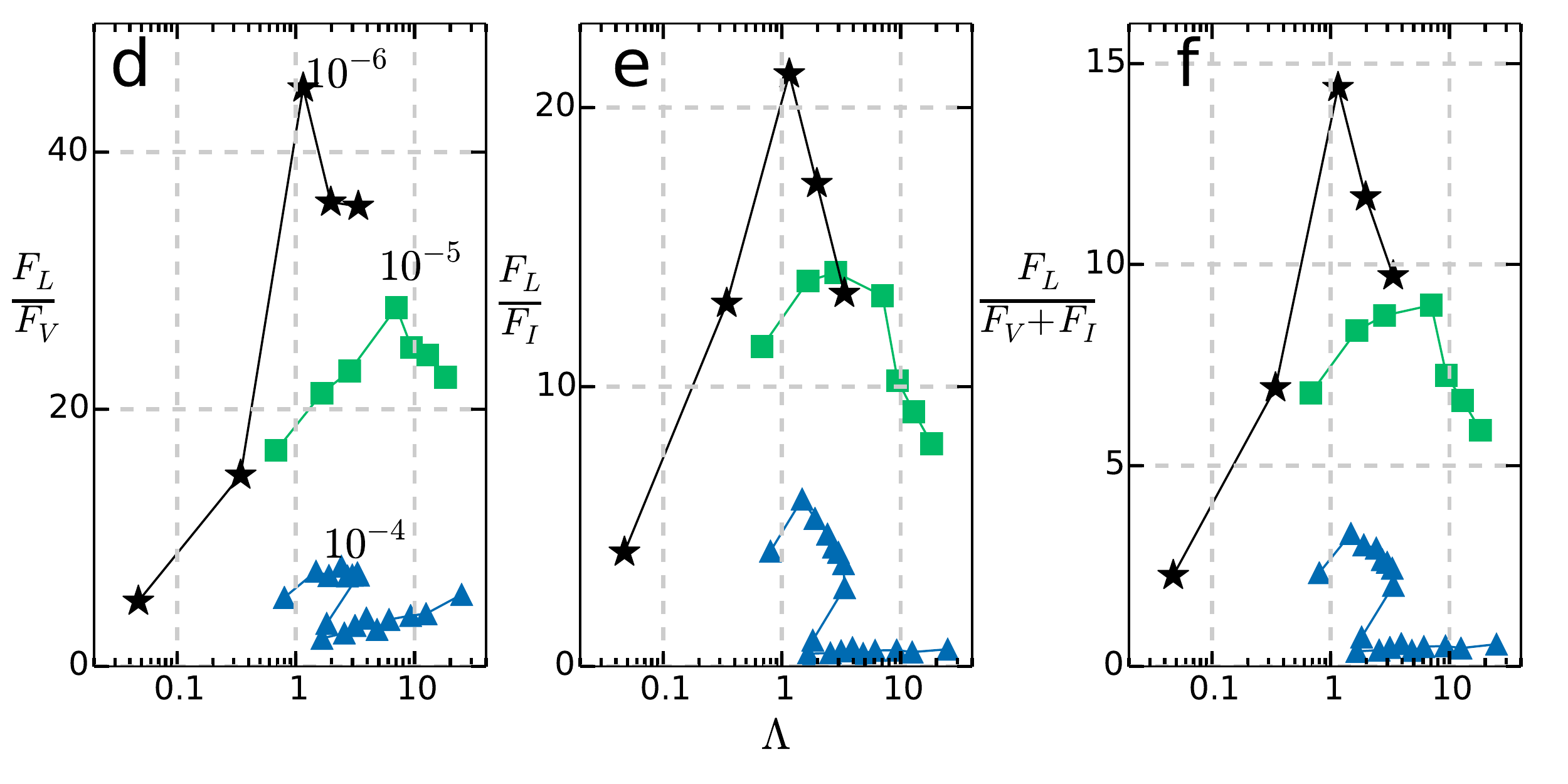}
\caption{Panels a, b, and c show the variation of the forces governing the dynamo simulations as a function of the convective supercriticality $Ra/Ra_c$. The $Ra_c$ values  assumed for $E=10^{-4},\,10^{-5},\,10^{-6}$ are $6.96\times 10^5$, $1.06\times 10^7$, $1.79\times 10^8$, respectively (Christensen \& Aubert \cite{christensen2006}). The magnitudes of Coriolis force and the pressure gradient force are similar for most Rayleigh numbers and the data points overlap. The legend describing the data in panels a, b, and c is shown at the top.  Panels d, e, and f show the behavior of various force ratios as a function of the dynamo generated Elsasser number $\Lambda$. The different colors in lower panels represent different Ekman numbers that are indicated in panel d. The curves connecting the $E=10^{-4}$ data points in the lower panels follow increasing $Ra$ trend. Therefore, as $E=10^{-4}$ dipolar dynamos become unstable at certain $\Lambda$, the curve turns back even though the $Ra$ increases.}
\label{fig:fig1}
\end{figure*}

\begin{figure*}
\centering
\includegraphics[width=1\linewidth]{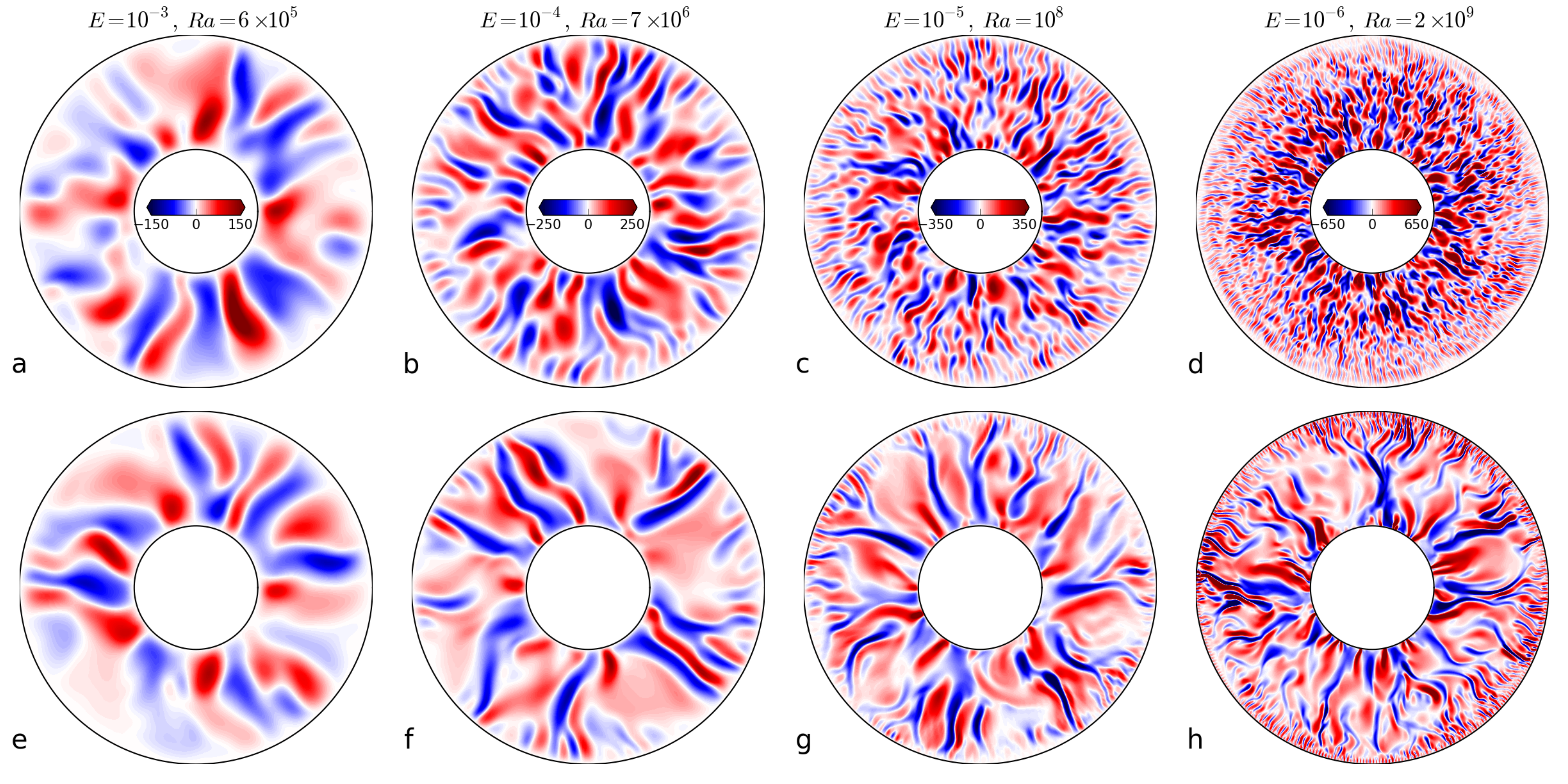}
\caption{Panels a, b, c, and d show the radial velocity, given in terms of the Reynolds number ($u\,D/\nu$, where $u$ is the local velocity), in the equatorial plane of the hydrodynamic simulations. Panels e, f, g, and h show the same for the corresponding magnetohydrodynamic cases. The Rayleigh number of all the cases shown is about 10 times $Ra_c$. The color maps are saturated at values lower than the extrema to highlight fainter structures.}
\label{fig:fig2}
\end{figure*}

\section{Results}
We begin our analysis by explicitly calculating the various forces involved in the system, namely, Coriolis force $F_C$, buoyancy (Archimedian) forces $F_A$, Lorentz force $F_L$, inertial force $F_I$, viscous force $F_V$, and pressure gradient force $F_P$. We compare the root-mean-square values of these forces, averaged in space and in time. Since our main goal is to compare the importance of various forces for the flow dynamics, care must be exercised in choosing the appropriate quantities. The spherically-symmetric component of any force is dynamically irrelevant; we thus exclude the harmonic order $m$=0 component from the force values. The PTC implies that the Coriolis force is largely compensated by the pressure gradient. For our purpose, we only concentrate on that part of $F_C$ that is not balanced by the pressure gradient force. Therefore, we consider $\mathbf{F}_C + \mathbf{F}_P$ rather than $\mathbf{F}_C$ alone. 

Since we employ no-slip boundary conditions, Ekman layers are formed at the boundaries \cite{greenspan1968}. Within these layers, the viscous force is dominant. Due to the larger viscosity, contemporary geodynamo simulations have much thicker Ekman boundary layers than those present in the Earth's core. This leads to a rather substantial contribution of the boundary layer viscous force to the total viscous force (e.g.~see \cite{stellmach2014,plumley2016}). To correct for this, we choose to exclude thin boundary layers, one below the outer boundary and one above the inner boundary, from the force calculation. The thickness of the excluded layers is 1, 2, 3\% of the shell thickness for $E=10^{-6}$, $10^{-5}$, $10^{-4}$, respectively. The chosen thickness of the layers is a rough estimate and the values are such that any larger value does not lead to further appreciable change in the bulk viscous force. For the sake of consistency, boundary layers are excluded from averaging procedure for all other force types as well. Sometimes it is argued that Ekman suction in the viscous boundary layer \cite{greenspan1968} plays an essential role for creating flow helicity as an important prerequisite for magnetic field generation \cite{davidson2015}. However, we note that geodynamo simulation with a stress-free boundary that lack Ekman suction show quite similar results compared to models with rigid boundaries \cite{yadav2013a, aubert2014}, hence viscous boundary layer effects do not seem to play an essential role.

The various forces calculated from the simulations are portrayed in Fig.~\ref{fig:fig1} (a,b,c) as a function of the convective supercriticality $Ra/Ra_c$ ($Ra_c$ is the critical $Ra$ where convection starts). First, notice that our choice of using ${\bf F}_C+{\bf F}_P$ rather than ${\bf F}_C$ makes a substantial difference since both $F_C$ and $F_P$ are very strong, however, they cancel each other to a large extent. This implies that to the zeroth-order the system is in a geostrophic state, where $F_C$ and $F_P$ are dominant. The first-order deviations are balanced by other weaker forces; these may be Lorentz, viscous, or inertial forces. One may call this state a `quasi-geostrophic' one \cite{calkins2015}. In the $E=10^{-4}$ simulations, the various forces remain comparable to each other to within an order of magnitude. This series of runs spans a large range of $Ra/Ra_c$, covering the transition from dipole-dominant dynamos to multipolar ones (occurring at around $Ra/Ra_c\approx 30$ for $E=10^{-4}$). With decreasing Ekman number the transition shifts to higher values of $Ra/Ra_c$ \cite{kutzner2002}, which are not reached in our simulations with $E\le 10^{-5}$. The latter all have a dominantly dipolar magnetic field. As convection becomes more turbulent, the inertial force eventually becomes the most dominant force  in our $E=10^{-4}$ simulations. For low convective supercriticalities ($Ra/Ra_c <$  10), ${F}_C$ and ${F}_A$ are comparable for all $E$. The Lorentz force  ${F}_L$ starts to match these two forces as $Ra$ increases. At $E=10^{-5}$ and more obviously at $E=10^{-6}$ a clear hierarchy of forces becomes apparent for $Ra/Ra_c \gtrsim 10$. Inertial and viscous forces are at least a factor of 10 weaker than the others. Lorentz, Archimedean and (uncompensated) Coriolis forces are very similar in amplitude and must balance each other, i.e. the bulk of the fluid is in a dynamical MAC state. We reiterate that since Coriolis and pressure forces are individually rather strong, the zeroth-order force balance is largely geostrophic and the notion of a MAC state in our simulations is a first-order effect.

We plot the ratio of ${F}_L$ and  ${F}_V$ as a function of the Elsasser number $\Lambda$ in Fig.~\ref{fig:fig1}d. In simulations with $E=10^{-4}$, as the dynamo generated field strength increases, the ratio $F_L/F_V$ reaches a maximum of about 8. Lowering $E$ to $10^{-5}$ and $10^{-6}$ increases this maximum ratio to about 30 and 45, respectively. The largest ratios between $F_L$ and $F_V$ is reached for cases with Elsasser numbers of order one. As shown in Fig.~\ref{fig:fig1}e, the ratio $F_L/F_I$ also follows the same qualitative trend as $F_L/F_V$. Note that a MAC state can be disturbed by the viscous force, however, with increasing flow turbulence, the inertial force can also do the same \cite{hughes2016}. Therefore, it is appropriate to compare Lorentz force and the sum of viscous and inertial force. As Fig.~\ref{fig:fig1}f shows, such a comparison provides a succinct way of highlighting the overall dominance of the Lorentz force. In this context, it is worth pointing out that assuming a higher magnetic Prandtl number may help to increase the strength of the magnetic field, and, in turn, its influence on the flow \cite{christensen2006, yadav2013a, dormy2016}. However, whether such an approach is justified or not remains to be tested.

The trends in the forces highlighted above have important consequences for the properties of convection. When a VAC balance holds in rapidly rotating convection, the characteristic flow length scale $l_u$ is proportional to $D\,E^{1/3}$, i.e length scales become smaller with decreasing $E$ \cite{jones2000b, king2013b, roberts2013}. As shown in Fig.~\ref{fig:fig2}(a,b,c,d), the convective structures in our hydrodynamic (HD) simulations do follow this trend qualitatively as $E$ decreases. On the other hand, in the MAC regime, $l_u$ is expected to be similar to the system size and to remain independent of $E$ \cite{jones2000a, starchenko2002, roberts2013}. For simulation with $E\ge10^{-4}$, both HD and dynamo cases have rather similar convective length scales (Fig.~\ref{fig:fig2}e,f). At $E=10^{-5}$, the dynamo case has a higher tendency for elongated structures in the radial direction  and fewer up- and down-wellings in azimuthal direction (Fig.~\ref{fig:fig2}g) as compared to the HD case (Fig.~\ref{fig:fig2}c). At $E=10^{-6}$, the dynamo case has significantly larger length scales (Fig.~\ref{fig:fig2}h) than the corresponding HD setup (Fig.~\ref{fig:fig2}d). This increased influence of the magnetic field is also reflected in the total magnetic energy which exceeds the total kinetic energy more and more as $E$ is decreased (Supplementary figure \ref{fig:figS1}). Another interesting feature in the $E=10^{-6}$ dynamo case is the presence of a layer of small scale convection near the outer boundary. This is caused by a relatively weaker Lorentz force in these regions (Supplementary figure  \ref{fig:figS2}). We conclude that hints of a MAC regime appear at $E=10^{-5}$ \cite{takahashi2012, teed2015} but this regime is  more prominent at $E=10^{-6}$. Furthermore, in a single system, there might be regions where a MAC state prevails while in some other regions it may not (also see \cite{sreenivasan2006, dormy2016}). 

\begin{figure*}
\centering
\includegraphics[width=0.8\linewidth]{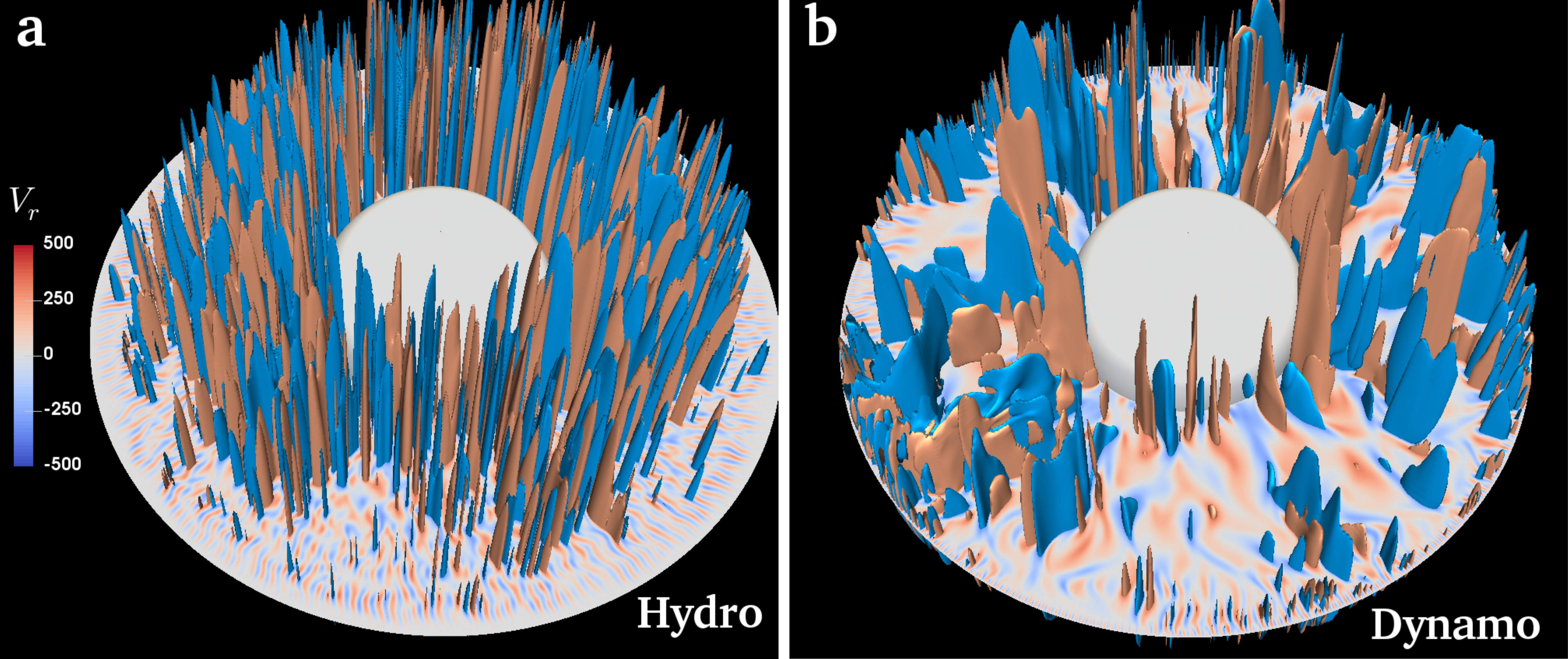}
\caption{Perspective view of a hydrodynamic (panel {\bf a}) and dynamo case (panel {\bf b}) with $E=10^{-6}$, $P_m=0.5$, $Ra=2\times10^9$. The radial velocity on the equatorial plane is given in terms of the Reynolds number. The blue and light orange contours represent radial velocity of -300 and 300, respectively.}
\label{fig:fig3}
\end{figure*}

\begin{figure}
\centering
\includegraphics[width=1\linewidth]{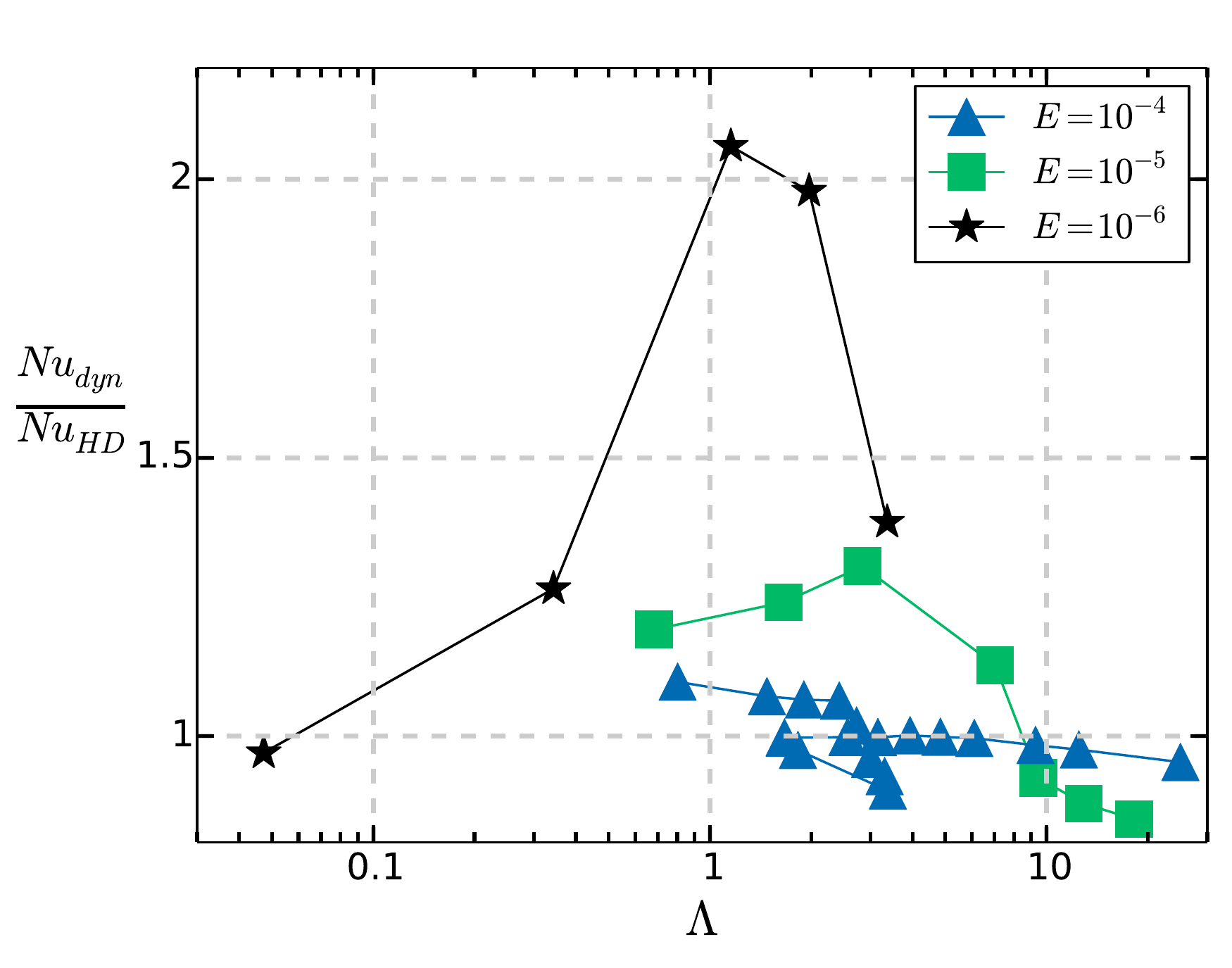}
\caption{Ratio of Nusselt number $Nu$ of dynamo and hydrodynamic cases (with otherwise same control parameters) as a function of the Elsasser number.} \label{fig:fig4}
\end{figure}

In Fig.~\ref{fig:fig3} we present the 3-dimensional morphology of the convection in the HD and in the dynamo case for the lowest viscosity simulation with the largest ratio of Lorentz force to viscous and inertial forces. The HD setup has small axially-aligned tube-like convection columns. In the dynamo case, however, the convection occurs in the form of thin sheets stretched in the cylindrically-radial direction.  It is also clear that as compared to the HD case the convective structures vary more along the rotation axis. Both features demonstrate the influence of the Lorentz forces on the convention morphology.

Another way to quantify the relaxed influence of the Proudman-Taylor condition in the dynamo cases is to analyze the total heat transferred from the bottom boundary to the top. This stems from the notion that rotation quenches the efficiency of convection by suppressing motions along the rotation axis \cite{greenspan1968}. Any relaxation of this constraint will lead to a gain in heat-transfer efficiency. We utilize the ratio of the Nusselt number $Nu$ (ratio of the total heat and the conductive heat transferred from the bottom to the top boundary) for dynamo and HD cases as a function of the dynamo generated average magnetic field strength (Fig.~\ref{fig:fig4}). At $E=10^{-4}$, the $Nu$-ratio remains close to unity, implying that the convective heat transport in dynamo and HD cases is similar.  At $E=10^{-5}$, the $Nu$-ratio peaks for $\Lambda\approx 3$ and reaches a value of about 1.3 \cite{yadav2016}. This enhancement of heat transport by the presence of a magnetic field is  more pronounced when we further decrease $E$ to $10^{-6}$. Here, the heat flow is doubled for $\Lambda\approx 1$. Comparing this figure with Fig.~\ref{fig:fig1}(d,e,f) highlights that the gain in the heat-transfer efficiency in the dynamo cases is largest  when the Lorentz force is maximally dominant over viscous and inertial forces.

\section{Discussion}

To summarize, we used a systematic parameter study to test the existence of a dynamical state in dynamo simulations  where magnetic forces play a crucial role together with Coriolis and buoyancy forces (MAC-state), as is expected to be present in the Earth's core. We lowered the viscosity to a small value, close to the limit allowed by today's computational resources, and found that Lorentz forces become equal in strength to (uncompensated) Coriolis and buoyancy forces and, for a limited range of Rayleigh numbers, far exceed viscous and inertial forces. This leads to large scale convection, substantial axial variation in the convection structures, and a 100\% increase in the heat-transfer efficiency as compared to the corresponding hydrodynamic setup. All of these features are expected theoretically \cite{roberts2013}. For higher viscosity values, the convection is much less affected by the magnetic field \cite{soderlund2012}. 

We note that in our simulations at the lowest Ekman number the Lorentz force is substantially smaller than the Coriolis force or the pressure force (taken individually). Hence, the state can be called quasi-geostrophic \cite{calkins2015}. Nonetheless, a completely geostrophic state is impossible and the essential question is what balances the residual Coriolis force. Since these are the Lorentz and Archimedean forces, with an insignificant role for viscosity and inertia, it is also justified to speak of a MAC-balance. We also note that although a MAC-balance is satisfied globally, this does not imply that the residual Coriolis force, Lorentz force and buoyancy force are pointwise of the same magnitude. For example, strong Lorentz forces seem to be rather  localized (see Supplementary figure \ref{fig:figS2}), as found in previous studies (e.g.~\cite{sreenivasan2006}). In regions where the Lorentz force is weak, the balance could be almost perfectly geostrophic or buoyancy alone could balance the residual Coriolis force.

Our results show some similarities with earlier studies done in a similar context. Larger scale convection in dynamo simulations compared to their HD counterparts has been reported in rotating convection in Cartesian geometry \cite{stellmach2004}; there, the dynamo simulation with $E=10^{-6}$ showed about 60\% increase in $Nu$. A recent laboratory experiment of rotating magnetoconvection (imposed magnetic field) in a cylinder also showed about 30\% increase in $Nu$ due to the presence of the magnetic field (at $E=4\times 10^{-6}$ and $\Lambda\approx 2$) \cite{king2015}. 

In the context of geodynamo simulations, studies at Ekman numbers comparable to the lowest value used in our study have been reported before. A substantial change in the convection length scale due to the dynamo generated magnetic field was found, but it only occurred  in cases with constant heat-flux boundary conditions \cite{sakuraba2009}. In contrast, we find the same enlargement of flow length scales also for fixed temperature conditions. Differences in the model setup and parameter values prevent us from elucidating the exact cause for these differences. Miyagoshi et al. \cite{miyagoshi2010, miyagoshi2011} also performed geodynamo simulations with $E\approx 10^{-6}$ (in our definition) and observed a ``dual convection" morphology where the deeper convecting regions had thin cylindrically-radial structures and the outer regions had very large scale spiraling features embedded into a prominent zonal flow. We also found such convection morphology at $E=10^{-6}$, in both hydrodynamic and dynamo simulations, but only at low Rayleigh numbers ($Ra/Ra_c <$ 10). Again, our simulations and these studies \cite{miyagoshi2010, miyagoshi2011} are significantly different in model details, for example they assumed gravity to drop sharply with radius whereas in our case it linearly increases from the inner to the outer boundary as it is appropriate for the Earth's core. A geodynamo simulation at the lowest Ekman number reached so far has been performed by Nataf and Schaeffer \cite{nataf2015} and shows rather small flow scales. Because hardly any details of the simulation are available it is difficult to assess the reasons. Possibly, strong driving could make inertial forces significant, leading to a compromised MAC state.

Our parameter study has shown that at an Ekman number of $10^{-6}$ a MAC-state, as is expected in the Earth's core, is very nearly reached, albeit only in a limited range of moderate Rayleigh numbers. As a consequence, the magnetic dipole dominates more strongly over higher multipoles at the outer boundary than it does in the geomagnetic field. Furthermore, the dipolar mode in the $E=10^{-6}$ simulation appears to be rather stable and does not show indications of reversals, unlike the geomagnetic field. In previous dynamo simulations, the onset of reversals has been associated with a growing influence of the inertial force at higher Rayleigh number \cite{sreenivasan2006,christensen2006}. We expect that pushing the Ekman number to even lower values would expand the range where a MAC-state exists towards more strongly supercritical values of the Rayleigh number \cite{christensen2010}, but this does not necessarily imply that inertia becomes significant. It remains an open question whether inertial effects are responsible for triggering reversals in the geodynamo (which would then not be in a pure MAC state), or if some other effects associated with a more strongly supercritical Rayleigh number play a role in reversals. Another challenge to tackle is the extreme value of the magnetic Prandtl number which is also fundamentally important for the geodynamo mechanism \cite{roberts2013}. In the Earth, $P_m$ is expected to be about $10^{-6}$, implying a large difference in the typical length scales of the velocity and the magnetic field (the latter varying on larger scales). To have a magnetic Reynolds number large enough to sustain a dynamo at low $P_m$, the convection must generate Reynolds number in excess of a million. In order to keep the system rotationally dominant {\em and} very turbulent one must inevitably decrease the Ekman number to much smaller values than what we could reach in this study. Therefore, a way forward in future is to strive for even lower Ekman numbers and lower magnetic Prandtl numbers to approach the conditions of the geodynamo. 

\begin{acknowledgments}
We thank the two anonymous referees for very constructive comments. Funding from NASA (through the {\em Chandra} grant GO4-15011X) and DFG (through SFB 963/A17)  is acknowledged. S.J.W. was supported by NASA contract NAS8-03060. Simulations were performed at GWDG and RZG.\end{acknowledgments}

\bibliography{cited}

\appendix

\begin{figure}
\centering
\includegraphics[width=1\linewidth]{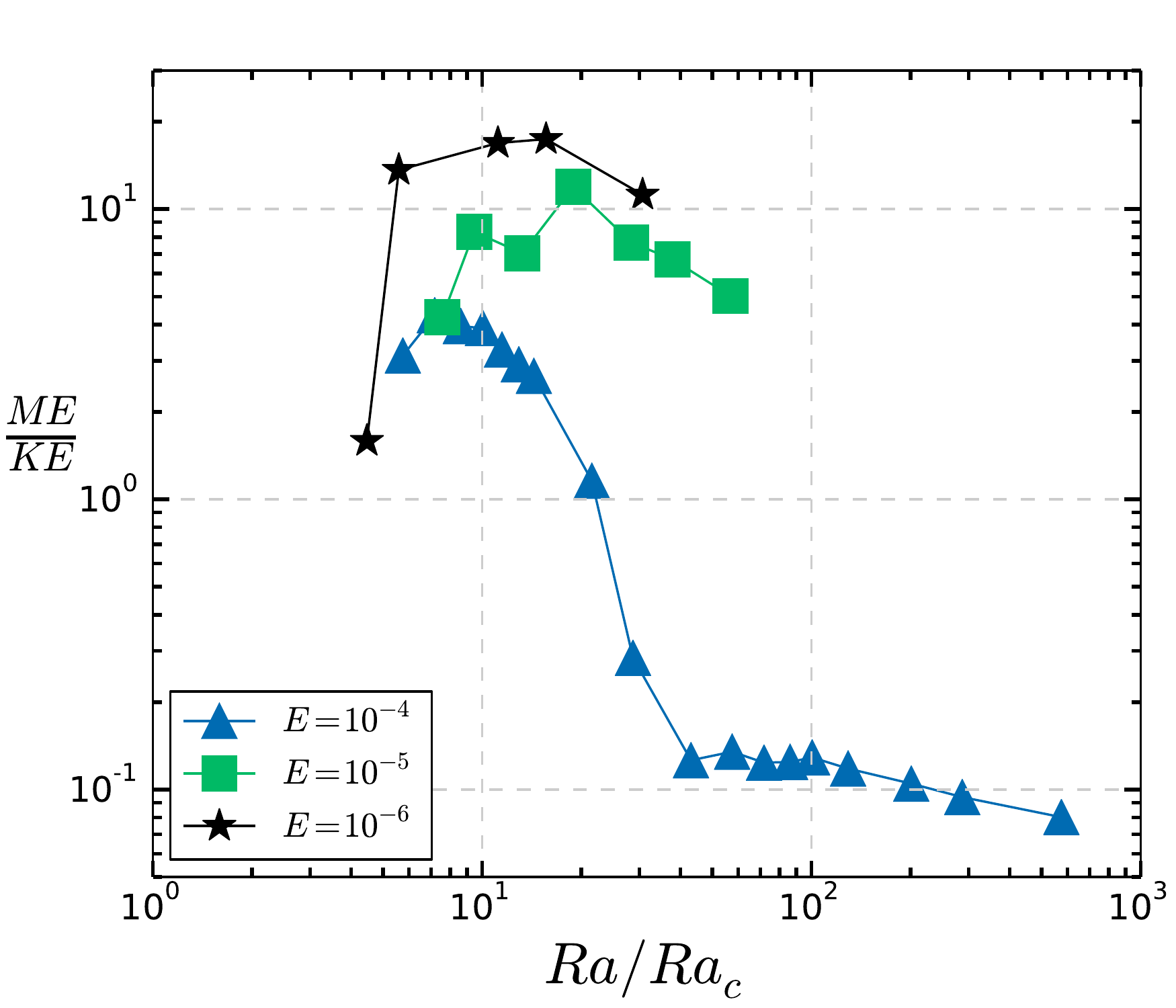}
\caption{({\bf Supplementary Figure}) Ratio of the total magnetic energy (ME) and the total kinetic energy (KE) as a function of the convective supercriticality.} \label{fig:figS1}
\end{figure}

\begin{figure}
\centering
\includegraphics[width=1\linewidth]{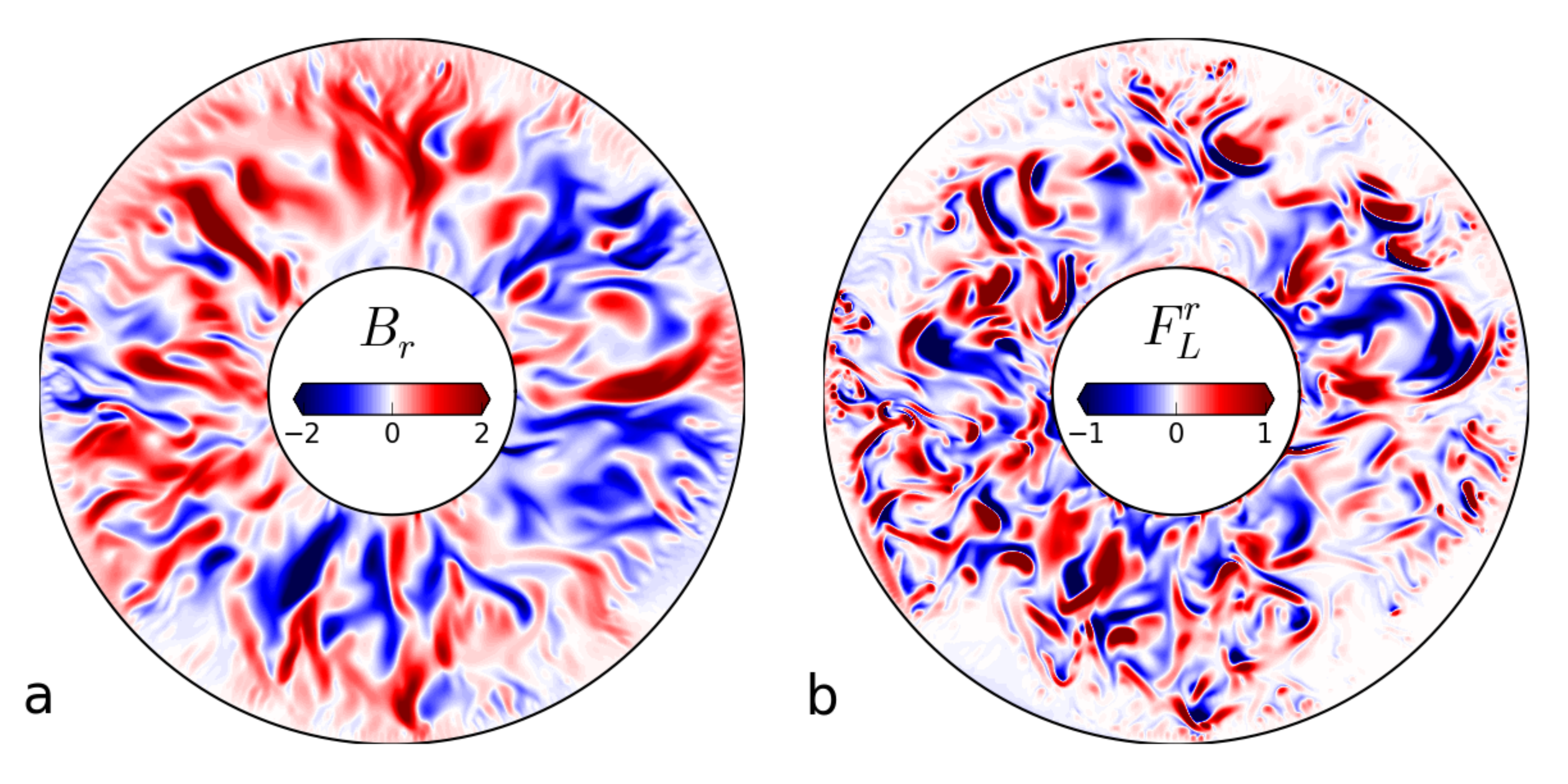}
\caption{({\bf Supplementary Figure}) The radial magnetic field in panel {\bf a} and the radial Lorentz force (down-scaled by $10^8$) in panel {\bf b} in the equatorial plane. The field strength is given in terms of the Elsasser number. The control parameters for this case are $E=10^{-6}$, $P_m =0.5$, $Ra=2\times 10^9$.} \label{fig:figS2}
\end{figure}

\end{document}